# Semantic Annotation and Search for Educational Resources Supporting Distance Learning

Mrs Nithya.C[#1], Mr K. Saravanan[*2]

[#] Student, [*]Assistant Professor
Department of Computer Science and Engineering
Regional Centre of Anna University
Tirunelveli (T.N) India

*Abstract*— **Multimedia educational resources play an important role in education, particularly for distance learning environments. With the rapid growth of the multimedia web, large numbers of education articles video resources are increasingly being created by several different organizations. It is crucial to explore, share, reuse, and link these educational resources for better e-learning experiences. Most of the video resources are currently annotated in an isolated way, which means that they lack semantic connections. Thus, providing the facilities for annotating these video resources is highly demanded. These facilities create the semantic connections among video resources and allow their metadata to be understood globally. Adopting Linked Data technology, this paper introduces a video annotation and browser platform with two online tools: Notitia and Sansu-Wolke. Notitia enables users to semantically annotate video resources using vocabularies defined in the Linked Data cloud. Sansu-Wolke allows users to browse semantically linked educational video resources with enhanced web information from different online resources. In the prototype development, the platform uses existing video resources for education articles. The result of the initial development demonstrates the benefits of applying Linked Data technology in the aspects of reusability, scalability, and extensibility.**

*Keywords*— **Linked Data, Semantic search, Cloud Applications, Web services, Semantic annotation, Ontology**

## I. INTRODUCTION

In the modern world e-learning activities are essential for distance learning in higher education. Digital video as one type of multimedia resource plays a vital role in distance learning. With the increased number of video resources being created , it is important to accurately describe the video content and enable searching of potential videos to enhance the quality and features of e-learning. It is critical to efficiently search for all related distributed educational video resources together to enhance the e-learning activities of the students. This paper adopts Semantic Web technology, more precisely; the Linked Data approach identified some primary challenges. Video resources should be described precisely, Descriptions of video resources should be accurate and machine-readable to support related search, Link the video resources to useful knowledgeable data from the web.

Video annotation ontology is designed by following Linked Data principles and reusing existing ontologies. It provides the foundation for annotating videos based on both time instance and duration in the video streams and more precise description details to be added to the video. A semantic video annotation tool (Notitia) is implemented for annotating and publishing education articles video resources based on the video annotation ontology. Notitia allows annotators to use domain specific vocabularies from the Linked Open Data cloud to describe the video resources. These annotations link the video resources to other web resources. A semantic-based video searching browser (Sansu-Wolke) is provided for searching videos. It generates links to further videos and education articles video resources from the Linked Open Data cloud and the web.

### A. Cloud Computing

Cloud computing is a model for enabling ubiquitous, convenient, on-demand network access to a shared pool of configurable computing resources (e.g., networks, servers, storage, applications, and services) that can be rapidly provisioned and released with minimal management effort or service provider interaction. This cloud model is composed of five essential characteristics, three service models, and four deployment models. The capability provided to the consumer is to deploy onto the cloud infrastructure consumer-created or acquired applications created using programming. Typically this is done on a pay-per-use or charge-per-use basis. A cloud infrastructure is the collection of hardware and software that enables the five essential characteristics of cloud computing. The cloud infrastructure can be viewed as containing both a physical layer and an abstraction layer. The physical layer consists of the hardware resources that are necessary to support the cloud services being provided, and typically includes server, storage and network components. The abstraction layer consists of the software deployed across the physical layer, which manifests the essential cloud characteristics. Conceptually the abstraction layer sits above the physical layer.

### B. Ontology Development – PROTÉGÉ 4.0

Ontology defines a common vocabulary for researchers who need to share information in a domain. It includes machine-interpretable definitions of basic concepts in the domain and relations among them.

**Sharing common understanding of the structure of information among people or software agents** is one of the more common goals in developing ontologies. For example, suppose several different Web sites contain medical information or provide medical e-commerce services. If these Web sites share and publish the same underlying ontology of the terms they all use, then computer agents can extract and aggregate information from these different sites. The agents can use this aggregated information to answer user queries or as input data to other applications.

**Enabling reuse of domain knowledge** was one of the driving forces behind recent surge in ontology research. For example, models for many different domains need to represent the notion of time. This representation includes the notions of time intervals, points in time, relative measures of time, and so on. If one group of researchers develops such ontology in detail, others can simply reuse it for their domains. Additionally, if we need to build a large ontology, we can integrate several existing ontologies describing portions of the large domain. We can also reuse a general ontology, such as the UNSPSC ontology, and extend it to describe our domain of interest.

**Making explicit domain assumptions** underlying an implementation makes it possible to change these assumptions easily if our knowledge about the domain changes. A hard-coding assumption about the world in programming- language code makes these assumptions not only hard to find and understand but also hard to change, in particular for someone without programming expertise. In addition, explicit specifications of domain knowledge are useful for new users who must learn what terms in the domain mean. Separating the domain knowledge from the operational knowledge is another common use of ontologies. We can describe a task of configuring a product from its components according to a required specification and implement a program that does this configuration independent of the products and components themselves.

**Analysing domain knowledge** is possible once a declarative specification of the terms is available. Formal analysis of terms is extremely valuable when both attempting to reuse existing ontologies and extending them. Often ontology of the domain is not a goal in itself. Developing ontology is akin to defining a set of data and their structure for other programs to be used. Problem-solving methods, domain-independent applications, and software agents use ontologies and knowledge bases built from ontologies as data.

### C. Linked Data Technology

Traditional video annotations using free-text keywords or predefined vocabularies are insufficient for a collaborative and multilingual environment. They do not properly handle the annotation issues, such as accuracy, disambiguation, completeness, and multi-linguality. For example, free-text keywords annotation easily fails on accuracy issues as they may contain spelling errors or be ambiguous. Furthermore, they are insufficient for a collaborative and multilingual environment. Our approach uses Linked Data to tackle the above issues in video annotations. It brings the following benefits. Each vocabulary is controlled and accurately defined in the Linked Data Cloud. It owns a unique URI to distinguish it from other vocabularies, so there are no conflicts between different vocabularies and meanings.

Linked data for improving student experience in searching e-learning resource persuades more and more people every day, because of the easy spreading web-based systems. Consequently, the web provides not only several data sources with useful and relevant information with e-learning purposes, but also information that is not easy to retrieve, and therefore wasted data. Sometimes, the information is inappropriate, and they must be filtered, in order to be relevant. Improving the way to exploit the web is an important step in the development of e-learning technology. The web would be a useful mechanism in the learning process if we take advantage of the information which is placed there. These improvements are related to the distribution of tasks: computers can be faster than humans in searching information with organized data. For humans, the search process on the web becomes a difficult task, and also very tedious, as a result of the large amount of information disposable to be consulted. For computers, large amount of data is not a problem.

## II. PROBLEM DEFINITION

Video resources should be described precisely. It is difficult to use only one general description to accurately tell the whole story of a video because one section of the video stream may have plenty of information (e.g., on historical figures and hidden events in the conversations) but some of them might not related to the main points of the video when it was created. Therefore, the normal paragraph-based description process is not good enough for annotating videos precisely. A more accurate description mechanism, based on the timeline of the video stream, is required.

The descriptions of the educational resources should be accurate and machine-understandable, to support related search functionality. Although a unified and controlled terminology can provide accurate and machine-understandable vocabularies, it is impossible to build such a unified terminology to satisfy different description requirements for different domains in practice. Link the video resources to useful knowledge data from the web. More and more knowledge and scientific data is published on the web by different education and educational organizations (e.g., Linked Open Data), and so it is useful to break the teaching resource boundaries between closed institutions and the Internet environment to provide richer learning materials to both educators and learners.

### III.  RELATED WORK

#### A. *A Lightweight Approach to Semantic Annotation of Research Papers*

A novel application of a semantic annotation system, named Cerno, to analyse research publications in electronic format. Specifically, we address the problem of providing automatic support for authors who need to deal with large volumes of research documents. To this end, we have developed Biblio, a user-friendly tool based on Cerno. The tool directs the user's attention to the most important elements of the papers and provides assistance by generating automatically a list of references and an annotated bibliography given a collection of published research articles. The tool performance has been evaluated on a set of papers and preliminary evaluation results are promising.

#### B. *Video Annotation through Search and Graph Reinforcement Mining*

Graph reinforcement method driven by a particular modality (e.g., visual) is used to determine the contribution of a similar document to the annotation target. The graph supplies possible annotations of a different modality (e.g., text) that can be mined for annotations of the target. Multimedia annotation algorithms can be said to be supervised or unsupervised based on whether it uses known training data. An annotation method can also be described as a computer vision approach if it builds word-specific models from low-level visual features, or a data mining approach if it mines correlations among annotations or propagates existing information. The graph reinforcement technique represents an inductive learning process that uses the weak predictions afforded by each similar video to create a stronger prediction of appropriate annotations for the set of videos.

#### C. *Semantic Video Search Using Natural Language Queries*

The indexing process assumes that the video annotations are made from a fixed set of vocabularies that change infrequently. Although this process can be efficient, the fixed set of vocabulary may introduce a gap between user's knowledge and indexed annotations, especially in the education environment, in which videos are often annotated by different groups of teachers or students, who may apply different annotation terms to the same video in the context of different courses and key points.
Ontology is a broader knowledge model with a reasoning mechanism that facilitates knowledge sharing on the semantic web. The knowledge representation language is used to create a set of terms and assumptions (axioms) about the meanings of the terms as well as to specify classes, properties and relationships between classes and objects in the domain.

#### D. *Multimodal Fusion for Video Search Re-ranking*

Videos are traditionally searched by syntactic matching mechanisms. Recently, with more videos being annotated or tagged in the Linked Data manner, educationists have begun to search videos in a more Semantic-Web oriented fashion. The two major approaches are the semantic indexing process and the natural language analysis process.

#### E. *Semantic Web for Content Based Video Retrieval*

The natural language analysis process focuses more on adding semantic tags to the user's search inputs. However, most of these approaches require machine learning mechanisms to assist dynamically adding tags. Hence, they restrict their applications to small and closed domains of discourse.

### IV. PROPOSED SYSTEM

Video annotation ontology is designed by following Linked Data principles and reusing existing ontology. It provides the foundation for annotating videos based on both time instance and duration in the video streams. This allows more precise description details to be added to the video.
 A semantic video annotation tool (Notitia) is implemented for annotating and publishing educational video resources based on the video annotation ontology. Notitia allows annotators to use domain specific vocabularies from the Linked Open Data cloud to describe the video resources. These annotations link the video resources to other web resources.
A semantic-based video searching browser (Sansu-Wolke) is provided for searching videos. It generates links to further videos and educational resources from the Linked Open Data cloud and the web.

- Annotations are accurate and free of spelling errors, ambiguity, and multi-linguality issues.
- The semantics of the annotations are process able by machine, which fosters the accuracy of searching and collecting related learning resources.
- The educational resources from different educational institutions are shared, reused, and semantically connected.

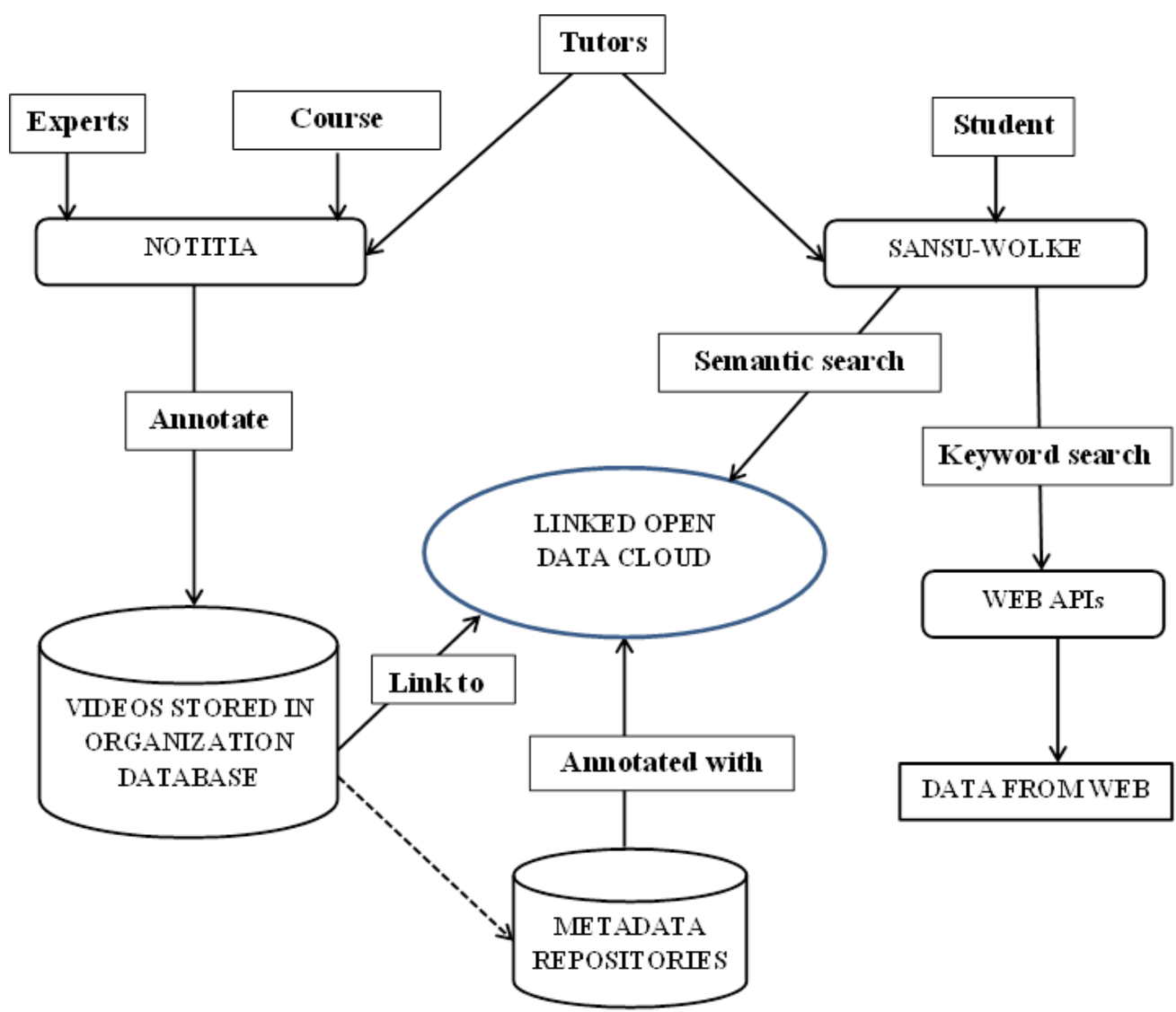

Fig. 1 Architecture of Linked data Technology
With Notitia and Sansu-Wolke

This process can be efficient; the fixed set of vocabulary may introduce a gap between user's knowledge and indexed annotations, especially in the educational environment, in which videos are often annotated by different groups of teachers or students, who may apply different annotation terms to the same video in the context of different courses and key points. The natural language analysis process focuses more on adding semantic tags to the user's search inputs. However, most of these approaches require machine learning mechanisms to assist dynamically adding tags. Hence, they restrict their applications to small and closed domains of discourse.

*A. Notitia – The Annotation Tool*

   The first tool is Notitia, which handles the input of annotations from users. Notitia provides a simple Web browser interface split into four main areas:

➢  a video player
➢  a list of videos, and existing annotations for the current video
➢  a set of controls for the video player, and input widgets to enter the annotations
➢  a set of panels to aid in finding suitable linked data annotations

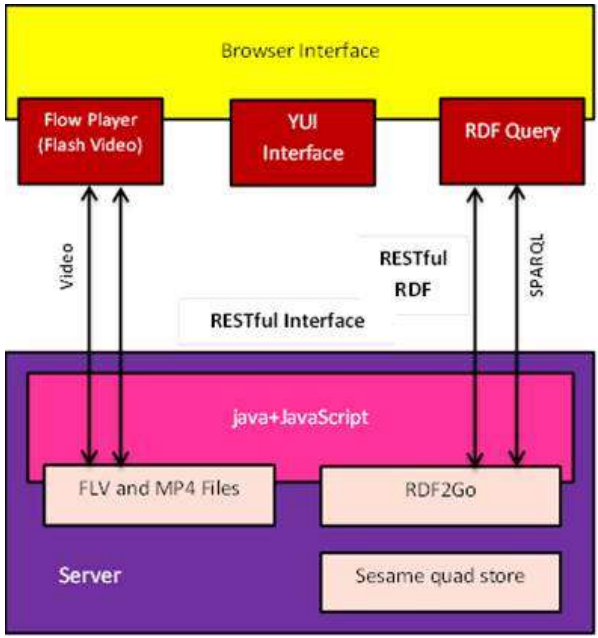

Fig. 2 Architecture of Notitia

In operation, the user first selects which of the videos to annotate. The videos of interest at the moment are transcoded at the request of the course team, and we serve them directly from the Notitia server, but the architecture supports using video served from anywhere on the Web.Having selected a video, the user will be presented with a timeline of any annotations previously made by herself and others: the user can skip to particular instants or durations noted in the annotations. Each vocabulary is controlled and accurately defined in the Linked Data Cloud. It owns a unique URI to distinguish it from other vocabularies, so there are no conflicts between different vocabularies and meanings.

Different vocabularies, which describe the same thing, are linked using the owl: sameAs property as an equation definition. Meanwhile, a number of semantic annotations are used to build the relationships between different vocabularies, such as rdfs: subclass of and rdfs. : see Also. Once a vocabulary is applied to an annotation, the related vocabularies are associated with the annotation. Therefore, the collaborative and multilingual issues are well addressed.
The most basic way to create an annotation is simply to pause the video at the appropriate point, enter duration if appropriate, and add a Semantic Web/Linked Data URI. This is sent to the server, and the annotation recorded.

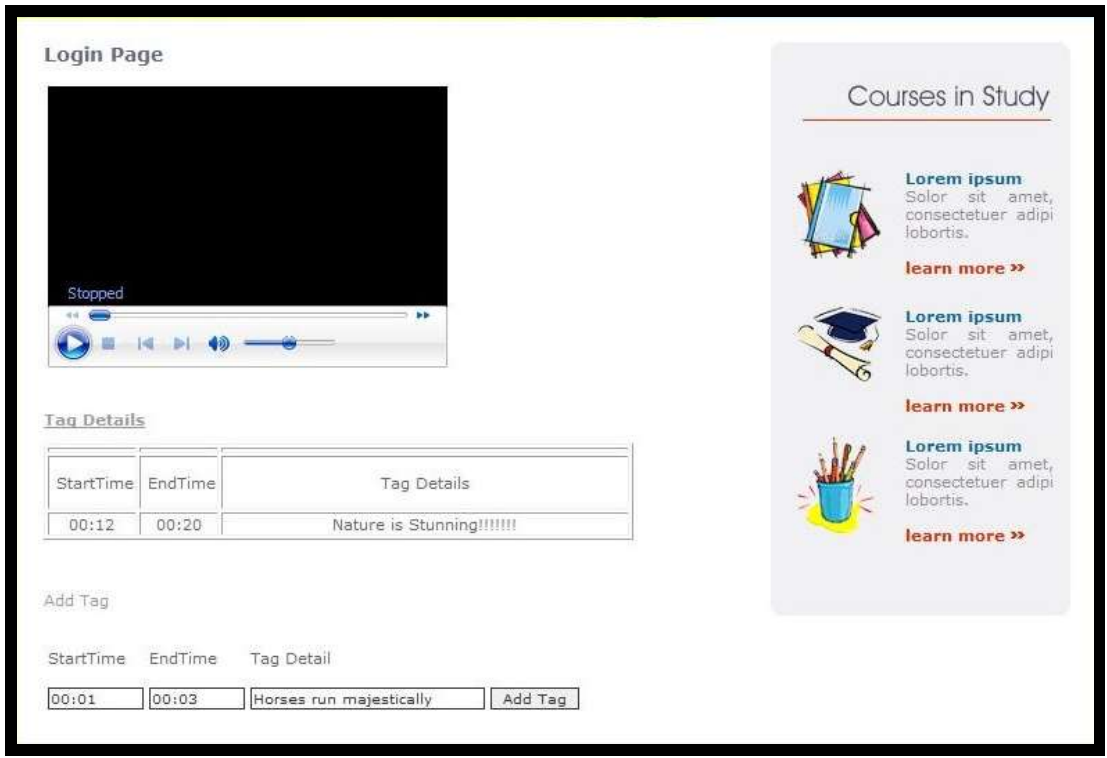

Fig. 3 Annotation of Video Resource

Annotations can be optionally marked as applying to something visible in the video, audible in the soundtrack, or to the conceptual subject matter of the video. Since finding appropriate URIs is non-trivial, the fourth part of the interface is dedicated to helping the user find them. We implemented the system in the java and jsp language, using Sesame as an RDF quad store and RDF2Go as an abstraction over the store. Notitia provides programmatic APIs in the form of a SPARQL end point for querying, and Restful interfaces for adding and removing annotations, and exploring existing ones. It is our intent that Notitia's RDF become part of the Linked Open Data cloud. In the client we use JavaScript with the Yahoo YUI library and the Query, and Flow Player for video playback.

### B. Sansu-Wolke – Semantic Search Browser

The second tool is Sansu-Wolke used to search the annotated videos, and explore related materials through Linked Data approach. The Sansu-Wolke architecture has two layers: the application layer, responsible for presentation and interaction, and the semantic data layer which finds and integrates data that has been published in the LOD cloud by different organizations, including annotation data from Notitia.

The Semantic Web data is obtained through Restful services and SPARQL endpoints. The application layer allows the user to query and navigate the video search results using the semantic data. Sansu-Wolke is designed to help the learner in navigating learning resources using Linked Data, so its first task is to connect the learner with appropriate concepts in the Semantic Web. Having found relevant conceptual URIs in the Linked Data cloud, Sansu-Wolke can then allow the user to navigate the links. Of course, Sansu-Wolke can query Notitia for videos that are tagged using the URIs of interest, but it can also find related video material through YouTube and Open Learn services. The user can be prompted with related concepts, Web documents, and other resources by Sansu-Wolke's analysis of neighbouring point LOD cloud, including Restful RDF services.

In addition to the Linked Data Services that are applied in the Notitia process, some other Linked Data Services and non-semantic services are used in Sansu-Wolke. The OU Linked Data that is currently under development and aims to extract and interlink previously available educational resources in various disconnected institutional repositories of the Open University and publish them into the Linked Open Data cloud and a semantic search engine, which crawls and collates the Semantic Web (including micro formats), and provides services such as keyword-based searching for linked data and accessing cached fragments of the Semantic Web. For example, when a user copies and pastes the learning content from lecture notes into the text field, all related knowledge concepts are listed, which enables the user to select further video searching activities. The Google map service is deployed for gathering the geo information about a place so that the user may click on the map to search related videos. The searching results do not only contain the OU educational video resources with their annotations but also include relevant learning resources about the videos and related videos from other services.

The Semantic Data Mining and Reasoning Layer has four different types of mining and reasoning processes: namely syntax parsing, document analysis, and annotation inferencing. The syntax parsing is the basic reasoning process to match syntax-based keywords to a URI identifier from the Linked Open Data Cloud. The syntax parsing process is triggered by the basic concept search functionality. The result of the parsing is a RDF description including the URI identifier. These syntax parsing results are the fundamental elements which perform the further video repository query and advanced reasoning.

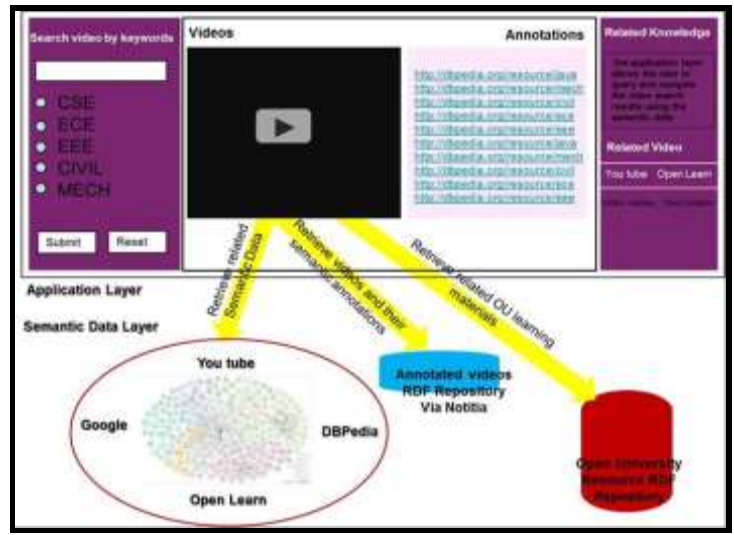

Fig. 4 Architecture of Sansu-Wolke

The document analysis process is used to analyse a document that is used to guide the study topic. Typical documents are lecture notes and online webpages. The analysis results are key learning points, knowledge, and concepts with their URI identifiers from the Linked Open Data cloud. These key points are matched to URI identifiers in DBpedia, Wikipedia, and Freebase for gaining further related educational resources.

The annotation inferencing process uses the tree-structure advantages of the ontology-based semantic annotations. By using the annotation reasoning process, the searching results are more accurate and widely covered. Although different video resource providers may use different Linked Data vocabularies to annotate their videos, they are linked together as search results through the Sansu-Wolke browser.

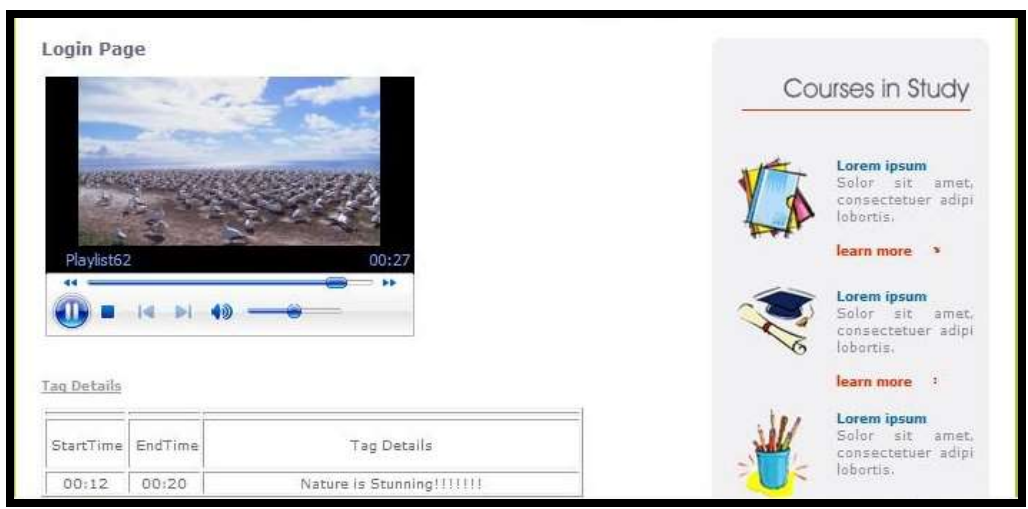

Fig. 5 Search Result of Annotated Video

## V. Evaluation and Result

The members all use eLearning systems on a daily basis for teaching distant learning students. In addition, three of them have experience of using textual tools for annotating the video learning materials. The students from ages 18- 28 are studying undergraduate courses at first year level. None of the students had any experience of using professional educational video searching tools before but often use Google or Yahoo when searching online. The evaluation process includes four steps:

**Demonstration:** we used one video as an example to show the annotation functionalities that use different Linked Data resources and web services. For students, we demonstrated how to use Sansu-Wolke with both basic concept search and advanced search

**Practice:** Two practice tasks are designed. One allows staff to annotate a certain video and the other allows student to search the videos related to the topics that the staff annotated.

**Evaluation:** We designed two sets of tasks for evaluating Notitia and Sansu-Wolke. Each set of tasks includes simple activities and more advanced activities such as using two different URIs to annotate one concept in the video or collaborative annotation. Each task has a 15-minute limit, and we monitored each user's time spent on each of the tasks.

**Feedback collection and analysis:** We used two evaluation questionnaires to collect feedback from users for Notitia and Sansu-Wolke, inquiring about the quality, performance and usability of the tools.

### A.  Notitia- The Annotation Tool

All members used an example "Java Tutorial" video. First, asking them to use free text or any references they would like to use to annotate the features in the video stream. Second, asking them to use You Tube or Google URI to identify the features in the video stream. Using the same video to try to find the concepts behind this video and annotate them. To do this, they should use the mode option to correctly identify if the event is clearly introduced in the video stream, conversation or the audio stream.

Using all different Linked Data suggestions from the suggestion panel to annotate the same video that is relevant to the "class concept" (at least three annotations).Trying to find another video that is related to the "class concept" class by searching for the Linked Data annotations. Asking people who sat next to each other to annotate the 2minute stream "Inheritance" video together,

- first, monitoring whether they both annotate the same items in the video;
- second, to check if they used the same Linked Data URIs for the same annotations;
- Third, if the same items are annotated by different Linked Data, to see if they can go through the link to find each other. Finally, to check if they can correct each other or get agreement to delete or keep the duplicated annotations.

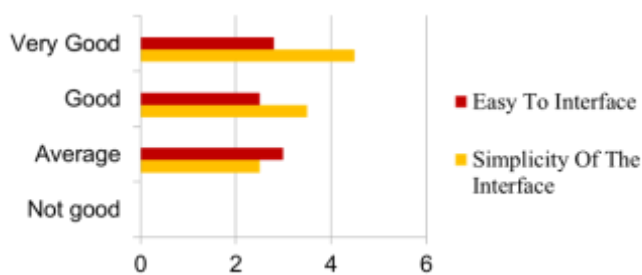

After the evaluation, we provided an Notitia evaluation questionnaire which consists of a rating for interface simplicity and usability, a rating for the quality and accuracy, identifying the most used annotation resources, identifying the most used annotation terms, and comments on using Linked Data technologies.

*B. Sansu-Wolke – Semantic Search Browser*

Students think the Sansu-Wolke can help their studies is the most interesting part of our evaluation task for SEG. The Sansu-Wolke evaluation tasks contains: Using the basic search functions to find as many as possible (no more related new video can be found) videos that related to topics and were stored in OU video repository. We examined the quantity of the found videos. Using the place search function and enter keyword to search for related videos. Go through all video resources from different video search providers to identify at least five videos that relate to topics.

Using the person search function and enter any person's name that you believe is related to class and identify at least two videos from different resources and two URIs that describe either the person we searched for related topics. Using the map search function to search for videos and information related to topic. Taking a particular text content from a lecture note that is used for the class to search the used annotation resources related and useful resources to prepare the class based on the highlights in the lecture note.

By monitoring how long it takes the students to identify all the related videos and useful information for a history lecture note (500 words), we found that most of the students can finish this task within 10 minutes. The videos or data voted most useful come from the OU linked open data set, Yahoo, NPTel and Google. The most important lesson learned from the evaluation at this stage is that students are more interested in the data that comes directly from the education-oriented services rather than social information websites such as YouTube.

The other parts of Sansu-Wolke evaluation questionnaire consist of the rating of the usability, quality, and accuracy of the tool. The major concern is the response time of some searches at runtime. As the Sansu-Wolke is a search tool that invokes different Linked Data services at the same time after search request is received, services' response time are different because of the quality of their own services and servers' runtime workload.

This is a trade-off between the quality and the accuracy. Since we invoke different Linked Data services at the runtime, the newest information and various data are found, which reflects the high accuracy satisfaction rate in the survey. Note that since both tools are still in prototype testing, there is still much work to be done to integrate them to the current OU distance learning systems and processes.

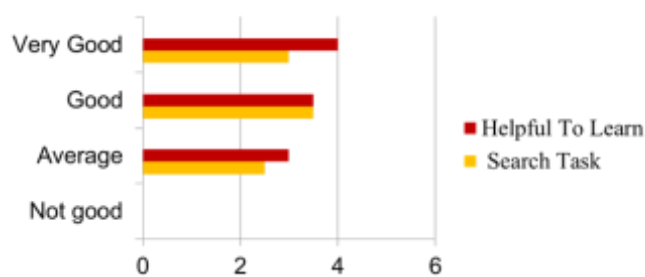

The limitation of our evaluation is that we cannot evaluate how much the Sansu-Wolke can improve the tuition without applying it on live course teaching and examination processes.

VI. CONCLUSION

This paper illustrated the Notitia and the Sansu-Wolke and no student thought platform that uses Linked Data technologies to semantically annotate and search educational video resources from the Open University video repository and link the videos to all other education resources on the web. In the semantic annotation process,

- Annotation of ontology is defined to support Linked Data annotations.
- Dynamic annotation URI suggestions are fully supported by integrating Linked Data Services into the Notitia interface;
- Collaborative functionalities are implemented to enhance the teamwork capability.

In the semantic search process, the search methods are based on the data retrieved through Linked Data Services and URIs, which links different resources together to enrich the original video search results. Sansu-Wolke shows that e-learning resources distributed across different education organizations can be linked together to provide more value-added information.

In education, collaboration has long been considered an effective approach for generate knowledge. Around the worldwide web, there are many contributions in terms of e-learning resources. These resources can solve the information needs of the people and can enhance the e-learning world. However, there are big challenges related to the resources information. One of them is the correct appropriation of these information, and other is the appropriate storage and use of these resources.

In this paper, we make use of the criteria selected by the students for creating an approach related to a collaborative e-learning environment, in which information provided by the LOD cloud diagram could be explored and use by different kind of e-learners, and also, we open the possibility to extend the linking open data community, giving the e-learners a place in which they can contribute, not only with new resources but also with qualifications of the information previously linked. This information would be used by other people in order to enlarge the collaboration regarding the knowledge around the worldwide web.

REFERENCES

[1] Hong Qing Yu, Carlos Pedrinaci, Stefan Dietze, and John Dominguez

[2] E. Allen and J. Seaman, "Class Differences Online Education in the United States," http://sloanconsortium.org/publications, 2010.

[3] J.W. Brackett, "Satellite-Based Distance Learning Using Digital Video and the Internet," IEEE Multimedia, vol. 5, no. 3, pp. 72-76, July-Sept. 1998.

[4] D. Wu, Y.-Y. Yeh and Y.-M. Chou, "Video Learning Object Extraction and Standardized Metadata," Proc. Int'l Conf. Computer Science and Software Eng., vol. 6, pp. 332-335, 2008.

[5] M. Hausenblas and M. Karnstedt, "Understanding Linked Open Data as a Web-Scale Database," Proc. Second Int'l Conf. Advances in Databases Knowledge and Data Applications (DBKDA), pp. 56-61, Apr. 2010.

[6] T. Berners-Lee, J. Hendler, and O. Lassila, "The Semantic Web," Scientific Am. Magazine, 2001.

[7] The Description Logic Handbook: Theory, Implementation, and Applications, F. Baader, D. Calvanese, D.L. McGuiness, D. Nardi Univ. of Denmark, master's thesis, 2010. and P.F. Patel-Schneider, eds. Cambridge Univ., 2003.

[8] Berners-Lee, "Linked Data," http://www.w3.org/Design Issues/LinkedData.html, 2006.

[9] C. Bizer, T. Heath, and T. Berners-Lee, "Linked Data—The Story So Far," Int'l J. Semantic Web and Information Systems, vol. 5, pp. 1- 22, 2009.

[10] E . Prud'hommeaux and A. Seaborne, "SPARQL Query Language for RDF," technical report, World Wide Web Consortium, Jan. University.2008.

[11] Heath, T ―Linked data - Connect Distributed Data across the Web . ‖ Retrieved June 24, 2011, from http://linkeddata.org/

[12] Klyne, G & Carroll, JJ 2004, ―Resource Description Framework (RDF): Concepts and Abstract Syntax. ‖ W3C Recommendation 10 February 2004. Retrieved May 22, 2011, from http://www.w3.org/TR/rdf-concepts/

[13] Kravari, K, Papatheodorou, K, Antoniou, G & Bassiliades Nick 2011, ― Reasoning and proofing services for semantic web agents, ‖ in Proceedings of the Twenty-Second International Joint Conference on Artificial Intelligence, Barcelona. Retrieved from http://www.aaai.org/ocs/index.php/IJCAI/IJCAI11/paper /viewFile/3164/3651

[14] McGuiness, DL & Harmelen, F van 2004, ―OWL Web Ontology Language. Overview.‖ W3C Recommendation 10 February 2004. Retrieved May 24, 2011, from http://www.w3.org/TR/2004/REC-owl-features-20040210/

[15] Prud'hommeaux, E & Seaborne, A 2008, ―SPARQL Query Language for RDF.‖ W3C Recommendation 15 January 2008. Retrieved May 24, 2011, from http://www.w3.org/TR/rdf-sparql-query/